\documentclass[sn-mathphys-num]{sn-jnl}

\usepackage{graphicx}%
\usepackage{multirow}%
\usepackage{amsmath,amssymb,amsfonts}%
\usepackage{amsthm}%
\usepackage{mathrsfs}%
\usepackage[title]{appendix}%
\usepackage{xcolor}%
\usepackage{textcomp}%
\usepackage{manyfoot}%
\usepackage{booktabs}%
\usepackage{algorithm}%
\usepackage{algorithmicx}%
\usepackage{algpseudocode}%
\usepackage{listings}%

\theoremstyle{thmstyleone}%
\theoremstyle{thmstyletwo}%

\theoremstyle{thmstylethree}%

\begin{document}

\title{Intriguing properties of transport at the microscales: Langevin equation approach}
\author{J. Spiechowicz and J. {\L}uczka}
\email{jakub.spiechowicz@us.edu.pl} 
\email{jerzy.luczka@us.edu.pl}
\equalcont{The  authors contributed equally to this work.}
\affil{Institute of Physics, University of Silesia, 40-007 Katowice, Poland}


%
%
\abstract{
We present a perspective of simple models of nonequilibrium directed transport described in terms of a Langevin equation formalism. We consider a Brownian particle under various circumstances and driven by thermal (equilibrium) and non-thermal (active) fluctuations. Three examples of startling behavior are unveiled: giant transport, multiple current reversal and negative mobility.
}


\maketitle



\section{Introduction}  
Transport of particles in micro and nano-world is strongly affected by fluctuations and random perturbations of various kind. In some regimes their role can be crucial. It can seem that their influence on the system is destructive and undesirable. But this is not the case at all. As an example, we can mention biological motors \cite{schilwa,motor} which operate in strongly fluctuating environment are good example of a constructive role of both equilibrium and non-equilibrium fluctuations. Biological motors possess many of the characteristics required to power molecular machines. They can generate force and torque, transport various cargos and are able to operate in a processive manner, i.e. they can move continuously along the specific substrates for distances of up to hundreds of steps (several microns) \cite{kay,processive}. 

In order to understand such real processes physicists use the method of idealization and construct simple models of particle transport at microscales and next, step by step, add subsequent elements which brings the theory closer to reality. We want to present this "step by step" procedure for a specific problem, namely, transport of a Brownian particle driven by both thermal equilibrium and nonequilibrium fluctuations. In this way we want to understand its generic properties, however, also reveal new ones which cannot be observed for macro-systems but are characteristic for the microscales \cite{resonance,reimann,hanggi,active,bressloff}. In doing so we can learn which components of the model are crucial for the emergence of these intriguing features and which elements are irrelevant.

Our survey is restricted to classical one-dimensional dynamics determined by the Langevin equation in the dimensionless form 
\begin{equation}
\label{L1}
		\ddot{x} + \gamma\dot{x} = -U'(x) + F(t) +  \sqrt{2\gamma D}\,\xi(t),  
\end{equation}
where $\gamma$ is the friction coefficient, $U(x)$ is the deterministic potential, $F(t)$ is an external force or non-equilibrium noise and $\xi(t)$ models thermal equilibrium fluctuations of zero-mean value $\langle \xi(t)\rangle=0$ and intensity $D$ which is proportional to temperature $T$. At first glance this equation seems to be very simple, however, in practice it is extremely versatile and exhibits impressive diversity of behavior which can successfully describe a number of important physical situations \cite{knoll1,knoll2,pnas,elife,commun,advances}. 
We want to stress that the above stingy model (1) does not pretend to describe biological motors but nevertheless  as we will demonstrate it can exhibit unintuitive transport behavior observed in real systems and therefore it may help to understand selected properties of much more complicated setups.  

The main characteristics of the particle transport is a non-zero averaged velocity in the long time regime, i.e. $\langle v \rangle \neq 0$ for $t \to \infty$. There are two distinct regimes of the dynamics in Eq. (\ref{L1}), namely the overdamped and underdamped one. In the first case the inertial term $\ddot{x}$ is neglected. In all cases presented below the long time stationary state of the considered system is a nonequilibrium state.
\begin{figure}[t]
	\centering
	\includegraphics[width=0.45\linewidth]{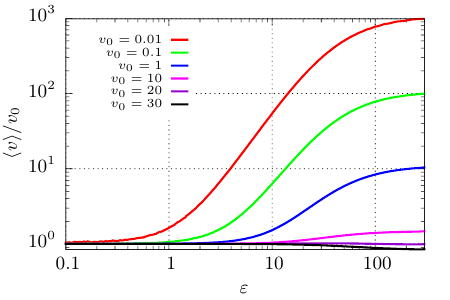}
	\includegraphics[width=0.45\linewidth]{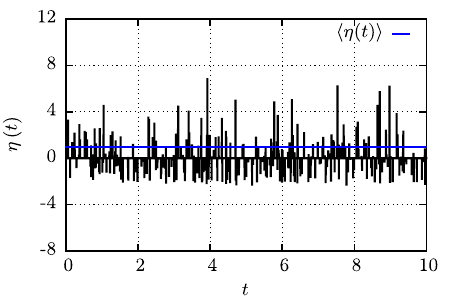}
	\caption{Giant transport amplification. The average velocity $\langle v \rangle$ of the Brownian particle dwelling in the symmetric potential $U(x) = \varepsilon \sin{x}$ and driven by Poissonian active fluctuations $\eta(t)$ vs the potential barrier height $\varepsilon$ for selected values of $\langle \eta(t) \rangle = v_0$.} 
	\label{fig1}
\end{figure}

\section{Giant Transport Amplification} 
The conventional way to transport the particle in a predefined direction is to apply a constant force $F$ pointing to that direction. Such a situation can be modeled by the simplified Eq. (\ref{L1}), reading
\begin{equation}
	\label{L2}
	 \gamma\dot{x} =  F +  \sqrt{2\gamma D}\,\xi(t).  
\end{equation}
We immediately see that $\langle v \rangle = F/\gamma \equiv v_0$. However, it is rather a trivial case. We can modify this model by replacing the deterministic load $F$ by the random force $\eta(t)$, namely,  
\begin{equation}
	\label{L3}
	 \gamma\dot{x} =  \eta(t) +  \sqrt{2\gamma D}\,\xi(t),   
\end{equation}
for which we postulate $\langle \eta(t) \rangle = F$ to be able to compare their influence. From Eq. (\ref{L3}) one gets $\langle v \rangle = F/\gamma = v_0$ as in the previous case (\ref{L2}). 
It follows that non-equilibrium noise of non-zero mean value $\langle \eta(t) \rangle = F$ has the same impact on $\langle v \rangle$ as the deterministic force $F$. Still, it is rather trivial. 

Now, let us locate the studied Brownian particle in a symmetric periodic substrate which in our theoretical modeling is translated into a unbiased spatially periodic potential $U(x)$. The dynamics of such a system is determined by the Langevin equations in the form
\begin{align}
	\label{L4}
	\gamma\dot{x} &= -U'(x) + F + \sqrt{2\gamma D}\,\xi(t), \\   
	\gamma\dot{x} &= -U'(x) + \eta(t) +  \sqrt{2\gamma D}\,\xi(t).    
	\end{align}
It could be expected that the directed velocity is reduced when the Brownian particle moves in the periodic potential because then it has to overcome the potential barriers. It is true for the deterministic force $F(t) = F$ so that $\langle v \rangle < v_0 $ \cite{risken}. However, when $F(t) = \eta(t)$ is a random force the directed transport in the periodic potential can be enhanced by many orders of magnitude, $\langle v \rangle \gg v_0$ \cite{bialas1,bialas2}. As an example of $\eta(t)$ we propose white Poissonian noise for which $\langle \eta(t) \rangle = F$. It has the form
\begin{equation}
	\eta(t) = \sum_{i=1}^{n(t)} z_i \delta(t - t_i),
\end{equation}
where $t_i$ are the arrival times of Poisson counting process $n(t)$ with parameter $\lambda$, which is the mean number of $\delta$-pulses per unit time. The amplitudes $\{z_i\}$ of $\delta$-kicks are statistically independent random variables sampled from the common probability distribution $\rho(z)$. 
Systems described by equations like Eq. (5) are an example of a model of nonequilibrium Markovian systems and their  deeper analysis is presented in Ref. \cite{vulpiani2}. 

In this case amplification of the directed velocity $\langle v \rangle/v_0$ can be many orders greater than one, see Fig. 1 where additionally an illustrative realization of the process $\eta(t)$ is included as well. Moreover, it shows that if the barrier height of $U(x)$ increases then $\langle v \rangle$ also grows. In consequence, dynamics within periodic structures can be many orders of magnitude greater than without it so that their existence is beneficial for maximizing the directed transport.  The detailed explanation of these properties is presented in Ref. \cite{bialas2}.

There are two characteristic time scales in this system: (i) the mean time $\tau_1=1/\lambda$  between $\delta$-kicks of Poissonian noise and (ii) the time $\tau_2$ the particle needs to move from the neighbourhood of the potential maximum to the vicinity of its minimum. The amplification of transport is maximal when $\tau_1 \approx \tau_2$. The resulting motion is synchronized: the particle is $\delta$-kicked and fall on one of the potential slopes, in the next time interval statistically there are no other $\delta$-spikes and it relaxes to a neighbouring minimum of the potential and this scenario repeats over and over again. However,  this mechanism is non-trivial and does not emerge for any distribution $\rho(z)$ of amplitudes $z_i$. The results in Fig. 1 are presented for the skew-normal distribution $\rho(z)$ \cite{azz}, which can be asymmetrical and both positive and negative amplitudes $z_i$ are possible \cite{bialas2}. \\
\begin{figure}[t]
	\centering
	\includegraphics[width=0.45\linewidth]{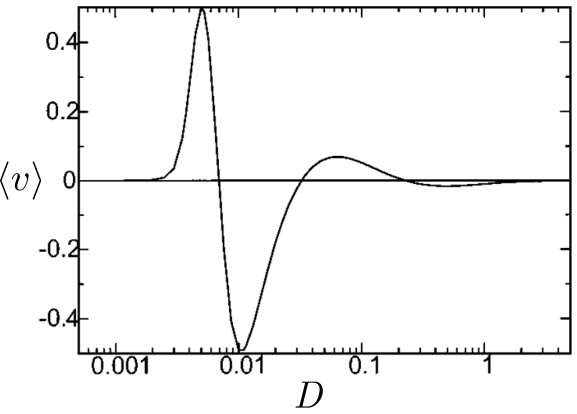}
	\includegraphics[width=0.5\linewidth]{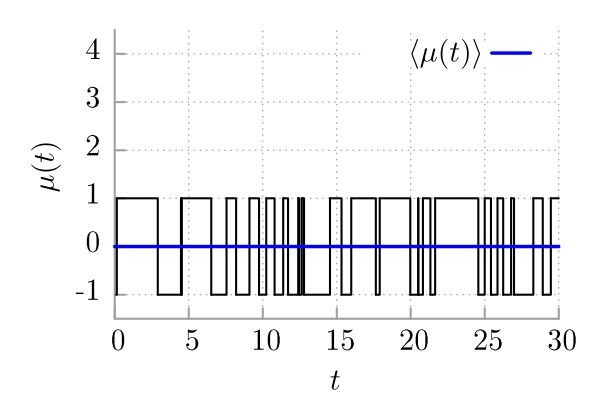}
	\caption{Multiple transport reversal. The average velocity $\langle v \rangle$ of the Brownian particle dwelling in the asymmetric piecewise linear potential $U(x)$ and driven by dichotomic fluctuations $\mu(t)$ vs temperature $D$ of the system.}
	\label{fig2}
\end{figure}

\section{Multiple Transport Reversal} 
An inversion of transport direction occurs when the average velocity of the particle changes its sign when a given parameter describing the system is altered \cite{kostur,knoll2,mateos,dykman}. It is a crucial problem for separation of particles at the (sub)-microscale. In order to present this issue we consider the Langevin equation in the form
\begin{equation}
	\label{L5}
	 \gamma\dot{x} =  -U'(x) + \mu(t) +  \sqrt{2\gamma D}\,\xi(t),   
\end{equation}
where $\mu(t)$ is a symmetric (zero-mean) dichotomic noise (a two-state Markov process), 
\begin{align}
	\mu(t) &= \{-a, a\}, \quad a>0, \\
	Pr(-a \to a) &= Pr(a \to -a) = \frac{1}{2\tau}, 
\end{align}
where $Pr(\pm a \to \mp a)$  is a probability per unit time (the rate) of jump from the state $\pm a$ to the state $\mp a$.
%
Because both $\xi(t)$ and $\mu(t)$ are symmetric, the directed motion of the particle is possible only when the symmetry of the spatial potential $U(x)$ is broken, e.g. it is in a ratchet form \cite{reimann,hanggi}. 

Such a case is depicted in Fig. 2, where an illustrative realization of dichotomous noise is included. The potential $U(x)$ is a piecewise linear function. 
Its definition is presented in Appendix 1 and depicted in Fig. 4.  
As it was shown in Ref. \cite{kostur}, within tailored parameter regimes, for $N$ maxima of $U(x)$ there are $N$ extrema of the average velocity as a function of temperature $D$ and $N-1$ temperatures which separate regimes of opposite directions of the particle transport. In the case shown in Fig. 2, the potential $U(x)$ has 4 maxima (see Fig. 4) and therefore there are 3 reversals of the velocity direction. The simple explanation of this behavior is not possible because there is a mixing of three forces. However, for very low thermal noise intensity $D$, the direction of velocity is determined by the set of equations 
\begin{equation}
	\label{L5}
	 \gamma\dot{x} =  -U'(x)  \pm a.  
\end{equation}
For different segments of the  piecewise linear potential, the particle moves with various  instant velocities determined by the slope $U'(x)$ of the potential and the noise value $\mu(t) =a$ or $\mu(t)=-a$. If $D$ increases the direction of velocity can change its sign and it is  determined by higher slope of the potential. It is the main mechanism for the velocity reversal.  This phenomenon can be detected when other parameters are changed as e.g. the friction coefficient $\gamma$ which,  via the Stokes relation, depends on linear size of the particle. In consequence, particles of different sizes can move in the opposite direction. 

Biological motors like kinesin and dynein move in the opposite direction on the same structure known as microtubule, i.e. the periodic substrate with broken reflection symmetry \cite{howard}. We do not claim that the above model describes movement of these motors but it shows that such a motion is possible even in a very simple setup and therefore from the physical point of view it can emerge also in more complex systems like biological cells.\\  
\begin{figure}[t]
	\centering
	\includegraphics[width=0.55\linewidth]{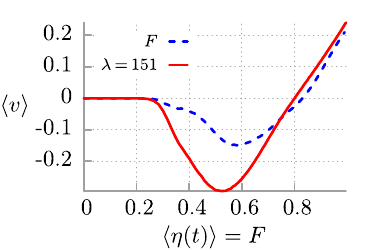}
	\caption{The negative mobility of the Brownian particle. The average velocity $\langle v \rangle$ of the Brownian particle dwelling in the symmetric rocking periodic potential $V(x) = U(x) - x \cos{(\omega t)}$ and driven by the constant force $F$ or Poissonian active fluctuations $\eta(t)$.}
	\label{fig3}
\end{figure}

\section{Absolute Negative Mobility} 
The next startling transport properties can be modeled by the equation  
\begin{equation}
	\label{L6}
	\ddot{x} + \gamma\dot{x} = -U'(x)  + a \cos{(\omega t)} + F +  \sqrt{2\gamma D}\,\xi(t).  
\end{equation}
Due to the presence of the external time periodic driving $a\cos{(\omega t)}$ the asymptotic state of this system is a time periodic nonequilibrium state \cite{jung1993} and therefore the directed velocity of the Brownian particle needs to be additionally averaged over its period, i.e.%
\begin{equation}
	\langle v \rangle = \lim_{t\to\infty} \frac{\omega}{2\pi} \int_{t}^{t+2\pi/\omega} {\mathbb E}[v(s)] \; ds,
\end{equation}
where $\mathbb E[v(t)]$ denotes the average of the actual velocity $v(t) = \dot x(t)$ over noise realizations and initial conditions. It is an example of a system in which the particle can move backwards against a constant force $F$. It means that $\langle v \rangle < 0$ when $F > 0$ and {\it vice versa}. This phenomenon is named the negative mobility or negative conductance \cite{machura2007, efficient, slapik, coex, eichhorn, vulpiani}. It is a minimal realistic model of the negative mobility which is tested experimentally \cite{ros, nagel}.
One can identify various regimes of the negative mobility: the absolute negative mobility around the zero bias $F \to 0$, the negative differential mobility $d\langle v \rangle/dF < 0$ and the emergence of negative nonlinear mobility remote from zero bias $F \neq 0$ \cite{machura2008}. Moreover, these anomalies can be induced by thermal fluctuations but can occur also in the deterministic ($D = 0$) chaotic and non-chaotic cases \cite{machura2007,strong,wisniewski}.  

In Fig. 3, the effect of negative mobility is illustrated. The characteristic feature is emergence of the interval in which the  constant force $F > 0$ can induce negative average velocity $\langle v \rangle < 0$. For comparison the influence of a random force in the form of white Poissonian noise $\eta(t)$ for which the amplitudes are distributed exponentially is shown as well.
We note that there exists an optimal value for the bias $\langle \eta(t) \rangle  \approx 0.58$ at which the average velocity $\langle v \rangle$ takes its minimal value. Most interesting is the fact that it is nearly two times lower than the one corresponding to the deterministic load $F$. It is the next example showing that the stochastic force (nonequilibrium fluctuations) can be more efficient than the deterministic perturbation. 

\section{Summary} 
In conclusion,  we presented three examples of intriguing properties of transport occurring at the microscales: giant amplification of directed velocity of the Brownian particle, multiple reversal of its direction and negative mobility. These illustrations show that in microworld equilibrium and nonequlibrium fluctuations can induce exotic features that are absent at macroscales. Moreover, contrary to common intuition, random perturbations may be beneficial for the system functioning as there are cases when they generate transport processes that are more pronounced than their counterparts caused by deterministic forces.

Physics of nonequilibrium phenomena has made milstone in sophisticated  experimental technics.  The future experimental setup   should be  used as a diagnostic tool to understand and  control of systems far from equilibrium. The observed scales of distances and times are now so small that allow to manipulate  molecules in simple fluids and  complex surroundings. E.g. many experiments are based on the single optical trap with either constant or 
time-dependent trap stiffness and single trajectories of particles are  monitored with high precision.

It is fascinating that such a simple model described by the Langevin equation (1) is able to capture a wide class of interesting phenomena. Moreover, despite a long history of investigation it still attracts a vibrant research activity. This little perspective article is by no means comprehensive in this regard. Other important closely related topics include the problem of diffusion \cite{anomalous,diffusion} and thermodynamics \cite{thermo} of this system. Time will tell how many more surprising effects will be discovered using this simple and elegant approach developed over 100 years ago by the founding fathers of modern physics.

\section*{Appendix 1}

The potential $U(x)=U(x+1)$ considered in Sec. 3 renders a piecewise linear function. Its visualization  is presented in Fig. 4. It is zero $U(x_0)=0$ at the points $x_0 \in \{0, 0.011, 0.04, 0.16, 1\}$ and its barrier heights are $U(x_i) \in \{1/8, 1/4, 1/2, 1\}$ at the points $x_i \in \{0.006, 0.025, 0.11, 0.44\}$, from the left to the right, respectively.  

\begin{figure}[t]
	\centering
	\includegraphics[width=0.45\linewidth]{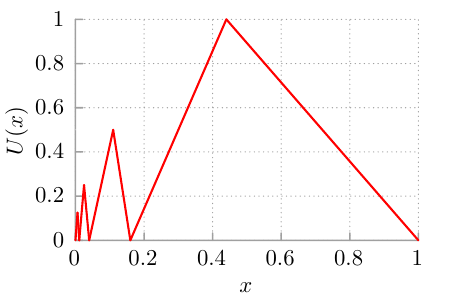}
	\caption{The piecewise linear periodic potential $U(x)$ considered in Sec. 3.}
	\label{fig4}
\end{figure}

\section*{Appendix 2}

We briefly sketch the numerical method used to simulate dynamics of the Brownian particle driven by non-thermal (active) fluctuations such as Eq. (5) or Eq. (7) with Poisson shot or dichotomic noise, respectively. The difference $s_k = t_k - t_{k-1}$ between subsequent times for the arrival of Poisson $\delta$-spike or switching of the dichotomic noise state is distributed exponentially $\psi(s) = \lambda \theta(s) \exp{(-\lambda s)}$, where $\theta(s)$ is the Heaviside step function. Therefore $s_k$ can be obtained by the transformation $s_k = -(1/\lambda)\ln{(1-y_k)}$ of independent random variables $y_k$ with uniform distribution over the interval $[0,1]$. The numerical procedure to calculate $x(t_k)$ from a given value $x(t_{k-1})$ is as follows: the random number $s_k$ is drawn from the distribution $\psi(s)$. Next, the ordinary differential equation corresponding to Eq. (5) or Eq. (7) is numerically integrated from $x(t_{k-1})$ to $x(t_k)$ and then the arrival of Poisson $\delta$-spike or the switching of the dichotomic noise state is taken into account. We refer the reader to Ref. \cite{numer} for the details of the algorithm.

\bmhead{Acknowledgements}
This work was supported by the Grant NCN 2022/45/B/ST3/02619 (J.S.)

\section*{Declarations}

\begin{itemize}
\item Funding:   Grant NCN 2022/45/B/ST3/02619 
\item Conflict of interest: The authors have no conflicts  to disclose
\item Ethics approval and consent to participate: not applicable 
\item Consent for publication: not applicable
\item Data availability:  Data sets generated during the current study are available from the corresponding author on reasonable request
\item Materials availability:  not applicable
\item Code availability : Codes are available from the corresponding author on reasonable request
\item Author contribution: all authors contributed equally to the paper
\end{itemize}

\end{document}